\documentclass[aps,prc,twocolumn,superscriptaddress,showpacs]{revtex4}

\usepackage{graphicx}
\usepackage{dcolumn}
\usepackage{bm}
\usepackage{ulem}
\usepackage{color}

\newcommand{\rgn}{($\gamma$,n)}
\newcommand{\rng}{(n,$\gamma$)}

\newcommand{\rdp}{(d,p)}

\newcommand{\sno}{$^{130}$Sn}
\newcommand{\sni}{$^{131}$Sn}
\newcommand{\snii}{$^{132}$Sn}
\newcommand{\sniii}{$^{133}$Sn}

\newcommand{\rpro}{r-process}

\begin{document}

\title{
  Direct capture in the $^{130}$Sn(n,$\gamma$)$^{131}$Sn and
  $^{132}$Sn(n,$\gamma$)$^{133}$Sn reactions under $r$-process conditions
}

\author{Peter Mohr}
\email{WidmaierMohr@t-online.de}
\affiliation{
Diakonie-Klinikum, D-74523 Schw\"abisch Hall, Germany}
\affiliation{
Institute of Nuclear Research (ATOMKI), H-4001 Debrecen, Hungary}

\date{\today}

\begin{abstract}
The cross sections of the $^{130}$Sn(n,$\gamma$)$^{131}$Sn and
$^{132}$Sn(n,$\gamma$)$^{133}$Sn reactions are calculated in the direct
capture model at low energies below 1.5\,MeV. Using recent data from (d,p)
transfer experiments on $^{130}$Sn and $^{132}$Sn, it is possible to avoid
global input parameters with their inherent uncertainties and to determine all
input to the direct capture model by local adjustments. The calculated direct
capture cross sections of $^{130}$Sn and $^{132}$Sn are almost identical and
have uncertainties of less than a factor of two. The stellar reaction rates $N_A
< \sigma v >$ show a slight increase with temperature. Finally an estimate for
the influence of low-lying resonances to the stellar reaction rates is given.
\end{abstract}

\pacs{24.50.+g,25.40.Lw,25.60.Tv,26.30.-k
}

\maketitle

Direct capture (DC) is expected to be the dominating reaction mechanism if the
level density in the compound nucleus is low. This is typically found for
light and/or neutron-rich nuclei, especially with magic proton or neutron
numbers, at low energies which is the relevant energy 
range for nuclear astrophysics. Direct neutron capture has been identified
experimentally for several stable targets (e.g.~$^{7}$Li \cite{Nag05},
$^{12}$C \cite{Kik98}, $^{16}$O \cite{Iga95}, $^{18}$O \cite{Mei96,Ohs08},
$^{22}$Ne \cite{Beer02,Tom03}, $^{26}$Mg \cite{Mohr98,Mohr99}, $^{48}$Ca
\cite{Beer96,Mohr97}), but it is obvious that neutron capture experiments are
practically impossible for short-living radioactive targets like \sno\ or
\snii . Thus, the determination of the DC cross section for unstable targets
has to rely on theoretical predictions.

The calculation of DC cross sections requires several
ingredients. First of all, the electromagnetic transition must be well
defined. This requires the transition energy $E_\gamma = E + S_n - E_x$ and thus
the neutron separation energy $S_n$ (or the masses of the target and residual
nucleus) and the excitation energy $E_x$ of the final state. In addition, spin
and parity $J^\pi$ of the final state and its spectroscopic factor $C^2 S$ are
essential ingredients for the calculation. Finally, the DC cross section
depends on the square of the overlap integral $\cal{I}$
\begin{equation}
{\cal{I}} = \int dr \, u(r) \, {\cal{O}}^{E1} \, \chi(r)
\label{eq:overlap}
\end{equation}
where ${\cal{O}}^{E1}$ is the electric dipole operator and $u(r)$ and $\chi(r)$
are the bound state wave function and scattering state wave function. These
wave functions are calculated from the two-body Schr\"odinger equation using
a simple nuclear potential without imaginary part because the damping of the
wave function in the entrance channel by the tiny DC cross sections is very
small \cite{Kra96}. The
present study is restricted to E1 transitions which are dominant in the DC
cross section whereas higher multipolarities like M1 or E2 are practically
negligible if dominant E1 transitions are allowed by the well-known
electromagnetic transition rules \cite{Xu12}. Further details of the DC model
can be found e.g.\ in \cite{Kra96,Beer96,Mohr98}.

The cross sections of the \sno \rng \sni\ and \snii \rng
\sniii\ reactions play an important role in \rpro\ nucleosynthesis.
In general, the influence of neutron capture cross sections on
\rpro\ nucleosynthesis is relatively small because under typical conditions an
equilibrium between \rng\ and \rgn\ reactions is reached. However, during
freeze-out the cross sections become important. This holds in particular for
the \sno \rng \sni\ reaction because of the larger neutron separation energy
$S_n = 5206 \pm 13$\,keV of \sni\ compared to the smaller $S_n = 2370 \pm
24$\,keV for \sniii\ (taken from the latest mass evaluation \cite{AMDC}). A
detailed study of the \rpro\ nucleosynthesis around $A \approx 130$ is given
in \cite{Sur09}. The particular importance of the \sno \rng \sni\ cross
section is 
highlighted in \cite{Beun09}, and the most important temperature range is
identified as $0.8 \le T_9 \le 1.3$ (where $T_9$ is the typical notation for
the temperature in $10^9$\,K). This corresponds to thermal energies
$70\,{\rm{keV}} \le kT \le 110$\,keV. Because of the missing Coulomb barrier
in neutron capture, the stellar reaction rate per mol and unit volume
${\cal{R}}(T) = N_A < \sigma v >$ (the usual short term ``reaction rate'' will
be used for ${\cal{R}}$ in the following) 
is mainly 
sensitive to the cross sections at energies around $E \approx kT$, and the
temperature dependence of ${\cal{R}}(T)$ is small (for pure $s$-wave capture
$\sigma \sim 1/v$ and ${\cal{R}}(T) = const.$).

Up to now, the DC cross sections of the \sno \rng \sni\ and \snii \rng
\sniii\ reactions have been calculated using global parametrizations of the
required input parameters \cite{Rau98,Chi08,Xu12}. It was found that the DC
cross section of the \sno \rng \sni\ reaction is very sensitive to the chosen
parameters. At 30\,keV a variation over three orders of magnitude is found
(see Fig.~9 of \cite{Rau98}). The recent
\rdp\ experiments on \sno\ \cite{Koz12} and \snii\ \cite{Jon11} allow for the
first time to completely avoid global parametrizations. Instead, locally
optimized parameters are used in this work for all ingredients of the DC
calculation to minimize the resulting uncertainties.

I start with the analysis of the DC cross section for the doubly-magic
\snii\ target nucleus. The bound state properties of the residual \sniii\ are
well-known from the \snii \rdp \sniii\ experiment \cite{Jon11} and are
summarized in Table \ref{tab:bound}. The spectroscopic factors are compatible
with unity (see Table I in \cite{Jon11}); thus, $C^2S \approx 1.0$ is adopted
in the following calculations. Such large spectroscopic factors are expected
for single-particle states above the doubly-magic \snii .
\begin{table}[tbh]
\caption{\label{tab:bound}
Properties of bound states in \sni\ and \sniii\ 
(from \cite{Koz12,Jon11,ENSDF,NDS131,NDS133})
and the considered E1 transitions.
}
\begin{center}
\begin{tabular}{rrrrr@{~~~~~}c@{$\rightarrow$}c}
\multicolumn{1}{c}{$J^\pi$}
& \multicolumn{1}{c}{$E_x$ (keV)}
& \multicolumn{1}{c}{$E$ (keV)}
& \multicolumn{1}{c}{~~~$C^2 S$~~~}
& \multicolumn{1}{c}{$V_0$ (MeV)}
& \multicolumn{2}{c}{$L_i \rightarrow L_f$} \\
\hline
\multicolumn{7}{c}{\sni } \\
$3/2^+$ &    0  & -5206 & 0.10 & -39.40 & 1,3 & 2 \\
$1/2^+$ &  332  & -4874 & 0.10 & -40.04 & 1   & 0 \\
$5/2^+$ & 1655  & -3551 & 0.10 & -36.82 & 1,3 & 2 \\
$7/2^-$ & 2628  & -2578 & 0.70 & -47.30 & 2,4 & 3 \\
$3/2^-$ & 3404  & -1802 & 0.70 & -46.97 & 0,2 & 1 \\
$1/2^-$ & 3986  & -1220 & 1.00 & -45.66 & 0,2 & 1 \\
$5/2^-$ & 4655  &  -551 & 0.75 & -43.70 & 2,4 & 3 \\
\hline
\multicolumn{7}{c}{\sniii } \\
$7/2^-$ &    0  & -2370 & $\approx 1.0$ & -46.51 & 2,4 & 3 \\
$3/2^-$ &  854  & -1516 & $\approx 1.0$ & -45.93 & 0,2 & 1 \\
$1/2^-$ & 1363  & -1007 & $\approx 1.0$ & -44.74 & 0,2 & 1 \\
$5/2^-$ & 2005  &  -365 & $\approx 1.0$ & -42.91 & 2,4 & 3 \\
\hline
\end{tabular}
\end{center}
\end{table}

The nuclear potential is taken as the sum of a central and a spin-orbit
potential
\begin{equation}
V(r) = -V_0 \, f(r) - V_{LS} \, \frac{{\rm{fm}}^2}{r} \, \frac{df}{dr} \,
\vec{L} \vec{S}
\label{eq:pot}
\end{equation}
with the central depth $V_0$, the spin-orbit strength $V_{LS}$ and the
Woods-Saxon geometry
\begin{equation}
f(r) = \Bigl[ 1 + \exp{(\frac{r-R}{a})} \Bigr]^{-1}
\label{eq:ws}
\end{equation}
with the radius parameter $R = R_0 \times A_T^{1/3}$, $R_0 = 1.25$\,fm, and $a
= 0.65$\,fm. 

In a first step the bound state wave functions $u(r)$ are calculated by
adjusting the depth $V_0$ of the central potential (with $V_{LS} = 0$) to the
energy $E < 0$ (see Table \ref{tab:bound}). With an additional spin-orbit
potential almost identical wave functions
can be obtained using $V_0 = 45.5$\,MeV (45.0\,MeV) and $V_{LS} = 18.6$\,MeV
(22.0\,MeV) for the bound $L=1$ ($L=3$) states.

The second step is the calculation of the scattering wave function
$\chi(r)$. The optical potential can be adjusted to experimental phase shifts
for all 
partial waves or to the scattering length for the $s$-wave. Unfortunately,
such data are not available for the unstable nuclei under study. As an
alternative, the potential strength can be adjusted to the energies of
single-particle states (as already done for the bound states above). For light
nuclei often a significant parity dependence for the potential depth $V_0$ is
found. However, with increasing mass this dependence decreases, and e.g.\ for
$^{49}$Ca (above the doubly-magic $^{48}$Ca) it is found that $V_0$
of the bound $L=1$ states (derived from the bound state energies) and $V_0$ of
the $s$-wave (derived from the scattering length) agree within about
1\,\%. (This result is independent of details of the geometry of the
potential; also for a folding potential the deviation is only about 1\,\%.)
\cite{Beer96}. Because of the minor difference of $V_0$ and $V_{LS}$ for the
$L=1$ and $L=3$ bound states, I adopt the average of $V_0 = 45.3$\,MeV and
$V_{LS} = 20.3$\,MeV for the calculation of the scattering wave functions
$\chi(r)$.

\begin{figure}
\includegraphics[width=\columnwidth,clip=]{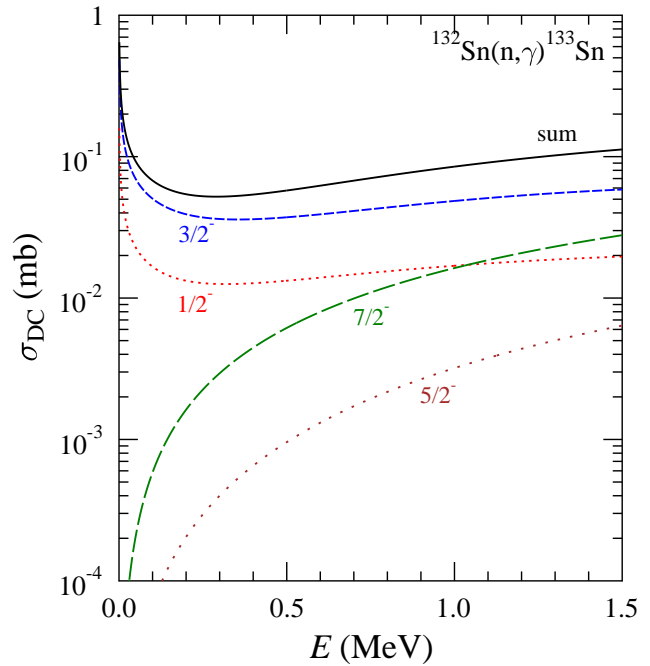}
\caption{
\label{fig:DC132}
(Color online)
DC cross section of the \snii \rng \sniii\ reaction. The contributions of the
bound states in Table \ref{tab:bound} are shown by colored dashed and dotted
lines. The full black line represents the sum over all bound states.
}
\end{figure}

Now all parameters for the calculation of the overlap integrals ${\cal{I}}$ in
Eq.~(\ref{eq:overlap}) are fixed by local adjustments to properties of
\sniii\ = \snii\ $\otimes$ n, and the DC cross sections can be calculated
without any further adjustments or parameters from global studies. The result
for the \snii \rng \sniii\ cross section is shown in Fig.~\ref{fig:DC132}. A
discussion of uncertainties will be given later.

Exactly the same procedure is repeated for the \sno \rng \sni\ reaction. The
bound state properties of the $L=1$ and $L=3$ bound states are taken from the
recent \sno \rdp \sni\ experiment \cite{Koz12}. Very similar to \sniii , no
fragmentation of the levels has been found in \sni\ which is somewhat
unexpected for the semi-magic \sno\ core (compared to the doubly-magic
\snii\ core in the previous case). The resulting average parameters $V_0 =
46.2$\,MeV and $V_{LS} = 21.1$\,MeV are derived from $V_0 = 46.6$\,MeV
(45.8\,MeV) and $V_{LS} = 20.3$\,MeV (21.9\,MeV) for the $L=1$ ($L=3$) bound
states. The potential parameters remain very close to the data for \sniii\ and
confirm the similarity of \sni\ and \sniii .

\begin{figure}
\includegraphics[width=\columnwidth,clip=]{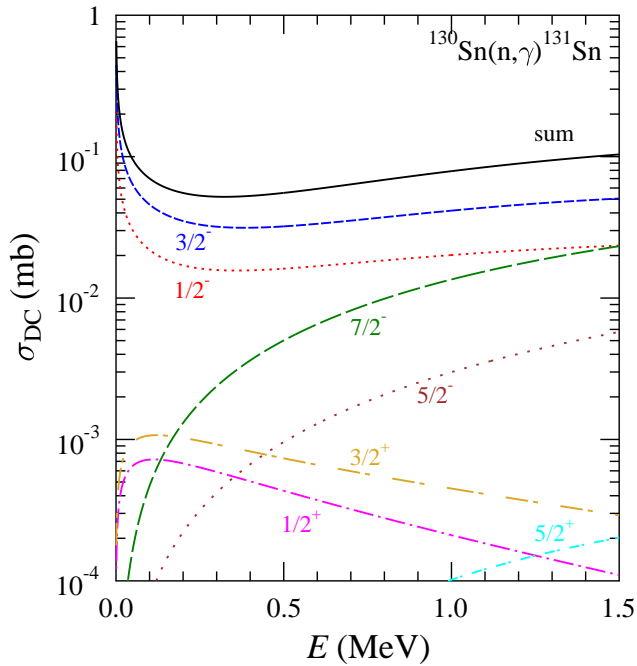}
\caption{
\label{fig:DC130}
(Color online)
DC cross section of the \sno \rng \sni\ reaction. The contributions of the
bound states in Table \ref{tab:bound} are shown by colored dashed and dotted
lines. The full black line represents the sum over all bound states.
}
\end{figure}

The bound states with even parity are characterized by a particle-hole
structure \cite{Koz12}. Thus, they have much smaller spectroscopic
factors. These states are not suited for a determination of the
potential depth $V_0$ which shows a broader spread. A spectroscopic factor of
$C^2 S = 0.1$ has been assumed for these states which is in agreement with the
upper limit of $\approx 0.3$ given in \cite{Koz12} but somewhat lower than the
average value of 0.347 for compiled spectroscopic factors
\cite{Xu12,Gor98}. The DC cross sections for the bound states with even parity
are much smaller than for the odd-parity bound states. The total DC cross
section (summed over all transitions) does not depend strongly on the assumed
value of $C^2 S = 0.1$ for the weak transitions to the bound states with
positive parity (see Fig.~\ref{fig:DC130}).

\begin{figure}
\includegraphics[width=\columnwidth,clip=]{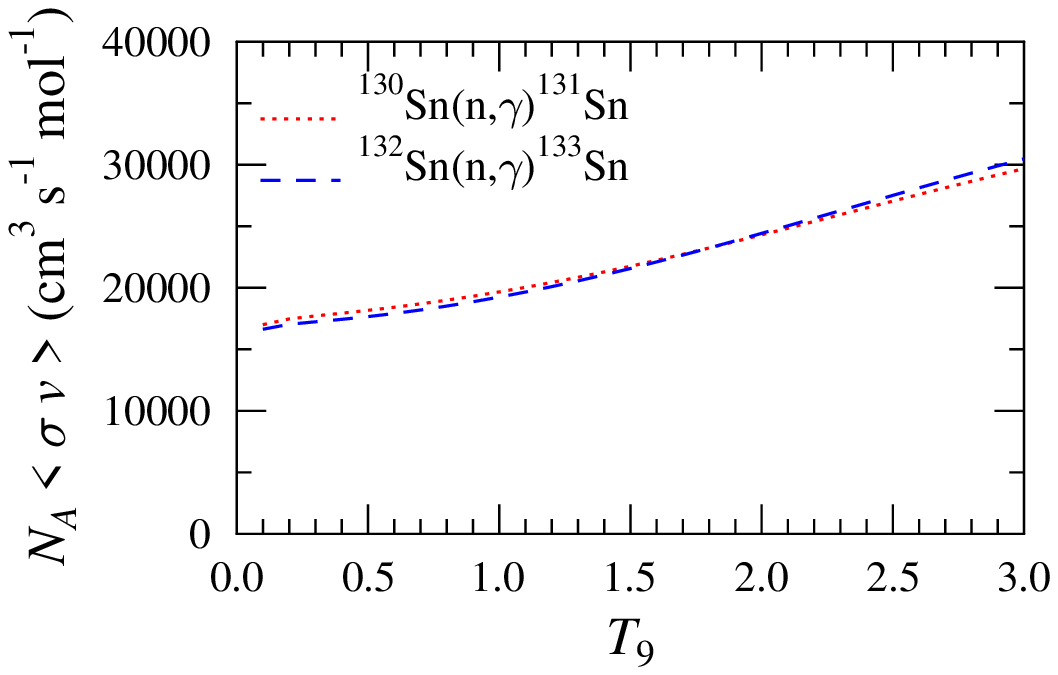}
\caption{
\label{fig:rate}
(Color online)
Stellar reaction rate ${\cal{R}}(T) = N_A < \sigma v >$ for the \sno \rng
\sni\ (red dotted) and \snii \rng \sniii\ (blue dashed) reactions.
}
\end{figure}

From the DC cross sections in Figs.~\ref{fig:DC132} and \ref{fig:DC130}
stellar reaction rates ${\cal{R}}(T) = N_A < \sigma v >$ can be calculated. 
Note that the laboratory reaction rate ${\cal{R}}_{\rm{lab}}$ and the stellar
reaction rate ${\cal{R}}^\ast$ are practically identical in the important
temperature range around $T_9 \approx 1$ \cite{Rau00}. The
reaction rates of both reactions under study are very similar and show a weak
temperature dependence (see Fig.~\ref{fig:rate}). The results can be simply
parametrized by a three-parameter parabolic fit
\begin{equation}
{\cal{R}}(T) = N_A < \sigma v > \approx \bigl( a_0 + a_1 T_9 + a_2 T_9^2
\bigr) \frac{{\rm{cm}}^3}{{\rm{s}}\ {\rm{mol}}}
\label{eq:rate}
\end{equation}
with $a_0 = 16811 \, (16321)$, $a_1 = 2291 \, (2236)$, and $a_2 = 700 \,
(870)$ for \sno\ (\snii ). The deviations of the fit are $1-2$\,\% over the
full temperature range under study.

Uncertainties of the DC cross sections are studied by a variation of the
different parameters of the calculation within reasonably estimated ranges and
by considering the experimental uncertainties of the bound state
properties. The uncertainty of the neutron separation energies $S_n$ and the
excitation energies $E_x$ lead typically to uncertainties for the transition
energy $E_\gamma$
of less than 10\,\%. Together with the $E_\gamma^3$ dependence of the
E1 transition strength a typical uncertainty of about $10-30$\,\% is found for
the various transitions under study. A variation of the potential geometry
(using a larger value of $R_0 = 1.4$\,fm instead of $R_0 = 1.25$\,fm) and
readjusting the potential depths leads to variations of the DC cross section
between $10-20$\,\%. A reduction of the potential depth $V_0$ by 3\,\% reduces
the DC cross section by about 15\,\%. The spectroscopic factors $C^2 S$ have
uncertainties of about 30\,\% which enter linearly into the DC
calculation. Combining all the above uncertainties of the order of
$10-30$\,\%, a total uncertainty below 50\,\% is a reasonable estimate for the
total DC cross section of the \sno \rng \sni\ and \snii \rng
\sniii\ reactions. 

For the \snii \rng \sniii\ reaction reaonable agreement with the three
predictions in \cite{Rau98} is found whereas the new result is lower by a
factor of slightly above 2 (slightly below 2) than the calculation in
\cite{Xu12} (\cite{Chi08}). The energy dependence of all calculations
\cite{Xu12,Chi08} is very similar because it is essentially defined by the
angular momenta in the entrance channel in combination with the
electromagnetic selection rules.

The obtained results for the \sno \rng \sni\ reaction are slightly below but
very close to the calculations shown in Fig.~4 of \cite{Koz12}. This is not
surprising because the same bound state properties ($J^\pi$ and $E_x$) are
used. The essential difference between this work and \cite{Koz12} is the
replacement of the global optical potential in \cite{Koz12} by the locally
optimized potential which reduces the uncertainties for the calculated
$\sigma_{DC}$. 

The new $\sigma_{DC}$ for \sno \rng \sni\ is about a factor of
two below the highest result by Rauscher {\it et al.}\ \cite{Rau98}. There are
two further calculations in \cite{Rau98} with much smaller cross sections
which result from the fact that some of the bound states in Table
\ref{tab:bound} are unbound in the corresponding calculations. However, the
dramatic reduction of the DC cross section in \cite{Rau98} is an artifact from
the separate treatment of the entrance and exit channels. If the $L=1$
bound states were indeed unbound, the $L=1$ strength would be located close
above threshold and show up as resonances in $\sigma_{DC}$ (and increase
$\sigma_{DC}$ via transitions to bound positive-parity states in \sni\ instead
of reducing $\sigma_{DC}$). This can be 
simulated by a reduction of the potential depth $V_0$, but is not taken into
account in \cite{Rau98} using a fixed potential in the entrance channel. E.g.,
using $V_0 = 
41.0$\,MeV (instead of 46.2\,MeV) leads to a strong $3/2^-$ resonance at about
73\,keV with a total width $\Gamma \approx 58$\,keV and a total cross section
of 4.2\,mb in the resonance maximum, i.e.\ a factor of about 50 higher than
the standard calculation 
shown in Fig.~\ref{fig:DC130}. The resulting stellar reaction rate ${\cal{R}}$
becomes temperature-dependent and would be a factor of $10-20$ higher than the
result in Fig.~\ref{fig:rate} because of this artificial $3/2^-$
resonance. However, such a strong resonance has been excluded by the transfer
data \cite{Koz12}.

Finally, predictions of the \sno \rng \sni\ and \snii \rng \sniii\ cross
sections from the statistical model have to be discussed briefly. As pointed
out e.g.\ in \cite{Rau00}, the statistical model is not applicable below $T_9
\approx 1.4$ for \snii\ and below $T_9 \approx 0.2$ for \sno\ because the
level density is too low. The limit for \sno\ may even be higher if one takes
into account that surprisingly low fragmentation of strength and very similar
properties of \sni\ and \sniii\ were found in the transfer experiments
\cite{Jon11,Koz12}. As a consequence, large deviations are found 
for predictions from the statistical model using different ingredients (for
details see Fig.~1 of \cite{Sur09} and discussion). Thus, a better estimate
for resonant contributions might be the procedure of lowering the potential
depth $V_0$ (as outlined above). 
From the spectroscopic factors in \cite{Koz12}
(see also Table \ref{tab:bound}) the missing $\approx 25$\,\% of the $L=1$ or
$L=3$ strengths may be located above threshold, but below the detection limit
of \cite{Koz12}. A resonance with full $3/2^-$ strength would lead to an
enhancement of the stellar reaction rate ${\cal{R}}$ by a factor of $10-20$;
thus, a weaker resonance with 25\,\% of the strength should enhance
${\cal{R}}$ not more than a factor of $2.5 - 5$ if located 
close above the threshold, and
the resonant enhancement is decreasing for higher-lying resonances. Such
an enhancement is only expected in the \sno \rng \sni\ reaction, but not for
the \snii \rng \sniii\ reaction because there are no bound states with
positive parity in \sniii\ \cite{ENSDF,NDS133}.

In summary, the direct capture cross section of the \sno \rng \sni\ and \snii
\rng \sniii\ reactions has been calculated using local parameters which could
be derived mainly from recent \rdp\ transfer experiments
\cite{Jon11,Koz12}. The DC cross sections of \sno\ and \snii\ are almost
identical and could be determined with relatively small uncertainties
of less than a factor of two. Additional resonant contributions may enhance
the stellar reaction rate by up to a factor of 5 for \sno\ depending on whether
the remaining $L=1$ and $L=3$ strength is located in a narrow energy window
close above threshold. Huge
enhancements of the reaction rate ${\cal{R}}$ of a factor of 10 or even 100 (as
discussed in \cite{Beun09}) are excluded by the present study.
\vspace{1mm}

\noindent
%
This work was supported by OTKA (NN83261).
%

\end{document}